# Estimating the prevalence of infectious diseases from under-reported age-dependent compulsorily notification databases

*Marcos Amaku[1], Marcelo Nascimento Burattini[1,2], Eleazar Chaib,[1] Francisco Antonio Bezerra Coutinho[1], David Greenhalgh[3], Luis Fernandez Lopez[1,4] and Eduardo Massad[1,5,\*]*

[1] LIM01-Hospital de Clínicas, Faculdade de Medicina Universidade de São Paulo, São Paulo, SP, Brazil

[2] Hospital São Paulo, Escola Paulista de Medicina, Universidade Federal de São Paulo, São Paulo, SP, Brazil

[3] Department of Mathematics and Statistics, The University of Strathclyde, Glasgow, Scotland, UK

[4] Center for Internet Augmented Research & Assessment, Florida International University, Miami, FL, USA

[5] London School of Hygiene and Tropical Medicine, London, UK

\*corresponding author (edmassad@usp.br)

**Abstract**

**Background:** National or local laws, norms or regulations (sometimes and in some countries) require medical providers to report notifiable diseases to public health authorities. Reporting, however, is almost always incomplete. This is due to a variety of reasons, ranging from not recognizing the diseased to failures in the technical or administrative steps leading to the final official register in the disease notification system. The reported fraction varies from 9% to 99% and is strongly associated with the disease being reported.

**Methods:** In this paper we propose a method to approximately estimate the full prevalence (and any other variable or parameter related to transmission intensity) of infectious diseases. The model assumes incomplete notification of incidence and allows the estimation of the non-notified number of infections and it is illustrated by the case of hepatitis C in Brazil. The method has the advantage that it can be corrected iteratively by comparing its findings with empirical results.

**Results:** The application of the model for the case of hepatitis C in Brazil resulted in a prevalence of notified cases that varied between 163,902 and 169,382 cases; a prevalence of non-notified cases that varied between 1,433,638 and 1,446,771; and a total prevalence of infections that varied between 1,597,540 and 1,616,153 cases.

**Conclusions:** We conclude that that the model proposed can be useful for estimation of the actual magnitude of endemic states of infectious diseases, particularly for those where the number of notified cases is only the tip of the iceberg. In addition, the method can be applied to other situations, such as the well known underreported incidence of criminality (for example rape), among others.

3**Background**

Compulsory notifiable diseases (CNDs) are those diseases that should be compulsorily reported to Health Authorities as soon as suspected by the attending professional [1] (Roush et al., 1999). The notified cases then enter a database from which, among other things, it is possible to know the incidence (new cases per age, sex, risk factor, geographic location, etc, per period of time) of the disease. The availability of such information allows health authorities, in principle, to monitor and to plan controlling the disease, for example providing early warning of possible outbreaks [2] (MMWR, 1998).

Almost 50 years ago, the World Health Organization's *International Health Regulations 1969* (*IHR/69*) already required disease reporting to the organization in order to help with its global surveillance and advisory role. These regulations were rather limited with a focus on reporting of only three main diseases: cholera, yellow fever and plague [3] (WHO, 1969).

The revised *International Health Regulations 2005* (*IHR/2005*), which entered into force on 15 June 2007, aims "to prevent, protect against, control and provide a public health response to the international spread of disease in ways that are commensurate with and restricted to public health risks, and which avoid unnecessary interference with international traffic and trade" [4] (WHO, 2005, p.4).

The IHR (2005) includes new demands to participant countries, such as: "(a) a scope not limited to any specific disease or manner of transmission, but covering "illness or medical condition, irrespective of origin or source, that presents or could present significant harm to humans"; (b) State Party obligations to develop certain minimum core public health capacities; (c) obligations on State Parties to notify WHO of events that may constitute a public health emergency of international concern according to defined criteria; (d) provisions authorizing WHO to take into consideration unofficial reports of public health events and to obtain verification from State Parties concerning such events; (e) procedures for the determination by the Director-General of a "public health emergency of international concern" and issuance of corresponding temporary recommendations, after taking into account the views of an Emergency Committee; (f) protection of the human rights of persons and travellers; and (g) the establishment of National IHR Focal Points and WHO IHR Contact Points for urgent communications between State Parties and the WHO" [4] (WHO, 2005, p.4).

In spite of international, national or local laws, norms or regulations requiring medical providers to report notifiable diseases to public health authorities, reporting is almost always incomplete [5-10] (Doyle, 2002; Gibbons et al., 2014; Keramarou and Evans, 2012; Rowe and Cowie, 2016; Gibney et al., 1991; Serra et al., 1999). This is due to a variety of reasons, ranging from not recognizing the diseased to failing in the technical and administrative steps leading to the final official report in the disease notification system. A recently published review [5] (Doyle et al., 2002), limited to published studies conducted in the United States between 1970 and 1999, quantitatively assessed infectious disease reporting completeness and found that reporting completeness varied from 9% to 99% and was strongly associated with the disease being reported. In another study [11] (Thacker et al., 1983), the mean reporting completeness for acquired immunodeficiency syndrome, sexually transmitted diseases, and tuberculosis as a group was significantly higher (79%) than for all other diseases combined (49%).

Of particular concern are those chronic, mainly asymptomatic, infectious diseases that allow infected individuals to live for years or even decades without being recognized as so. These diseases can represent a heavy burden to the affected populations and pose significant risk to the international community. Perhaps the most dramatic examples of the latter include human immunodeficiency (HIV) and hepatitis C (HCV) viruses pandemics. In fact, these two infections have been labeled by WHO as the epidemics of the XXth and XXIth centuries, respectively [9], [12] (Mann, 1996; Gibney et al., 2016).

One critical consequence of under-notification of such diseases is the fact that their prevalence estimates are frequently way underestimated, leading to miscalculation of their actual burden and making control efforts suboptimal [6] (Gibbons et al., 2014).

In a previous paper [13] (Amaku et al., 2016) we assumed that the infection (HCV) was in steady-state. Then we proposed two methods to give a first rough estimate of the actual number of HCV infected individuals (prevalence) taking into account the yearly notification rate of newly reported infections (incidence of notification) and the size of the liver transplantation waiting list (LTWL) of patients with liver failure due to chronic

HCV infection [14] (Chaib et al., 2014). Both approaches, when applied to the Brazilian HCV situation converged to the same results, that is, the methods proposed reproduce both the prevalence of reported cases and the LTWL with reasonable accuracy. In that paper we show how to calculate the prevalence of people living with HCV in Brazil, which resulted in a value up to 8 times higher than the official reported number of cases [13] (Amaku et al., 2016).

The present paper is an improvement of those techniques because, unlike in the previous paper mentioned above, now we do not assume steady state. Unfortunately, given the short period of time with data available (hepatitis notification became compulsory in Brazil only in 1999 [15] (MHB, 2000)), it cannot give more precise information on HCV prevalence than the one already provided by our previous study, but it illustrates the techniques that allow the prevalence estimation based on age and time of previous notifications, and that can be applied to any notifiable disease.

This paper is organized as follows. In Section 2, we describe the variables that can be extracted from the governmental notification database, by using two mathematical models. In Section 2.1 we describe a continuous model, that is a model where the variables are assumed to be continuous functions of time and age. In Section 2.2 we describe a discrete model, in which the variables are assumed to have only integer values in time and age.

**Methods**
**Formalism**
**Continuous time and age model**

Assume we have an SIR type infection and let $S(a,t)da$, $I(a,t)da$ and $R(a,t)da$ be the number of individuals with age between $a$ and $a+da$ at time $t$ that are susceptible, infected and removed (or recovered), respectively. In addition, as mentioned in the Introduction, Public Health authorities demand that some diseases be compulsorily notifiable, that is they publish the number of diagnosed individuals per time unit for each age interval (incidence) in public databases. Therefore, we can divide the prevalence of infected individuals into two classes: notified individuals, denoted $I^N(a,t)da$, and non-notified individuals, denoted $I^{NN}(a,t)da$.

Let $\lambda(a,t)$ be the so-called age and time-dependent force-of-infection (incidence density). Then:

$$\lambda(a,t)S(a,t)dadt \qquad (1)$$

is the number of susceptible individuals who get the infection when aged between $a$ and $a+da$ during the time interval $dt$. Standard arguments allow us to write the following system of partial differential equations, known as Trucco-Von Foester equations in the literature (Trucco, 1965):

$$\frac{\partial S(a,t)}{\partial t} + \frac{\partial S(a,t)}{\partial a} = -\lambda(a,t)S(a,t) - \mu(a,t)S(a,t),$$

$$\frac{\partial I^{NN}(a,t)}{\partial t} + \frac{\partial I^{NN}(a,t)}{\partial a} = \lambda(a,t)S(a,t) \qquad (2)$$
$$- \big(\mu(a,t) + \alpha^{NN}(a,t) + \gamma^{NN}(a,t)\big)I^{NN}(a,t) - \kappa(a,t)I^{NN}(a,t),$$

$$\frac{\partial I^N(a,t)}{\partial t} + \frac{\partial I^N(a,t)}{\partial a} = \kappa(a,t)I^{NN}(a,t) - \big(\mu(a,t) + \alpha^N(a,t) + \gamma^N(a,t)\big)I^N(a,t),$$

$$\frac{\partial R(a,t)}{\partial t} + \frac{\partial R(a,t)}{\partial a} = \gamma^{NN}(a,t)I^{NN}(a,t) + \gamma^N(a,t)I^N(a,t) - \mu(a,t)R(a,t),$$

where the meaning of the parameters is described in Table 1.

**Table 1. Parameters used in model (2).**

| Parameter | Meaning |
|---|---|
| $\lambda(a,t)$ | Force of Infection |
| $\mu(a,t)$ | Natural Mortality Rate |
| $\alpha^{NN}(a,t)$ | Disease-induced Mortality Rate for non-notified individuals* |
| $\alpha^N(a,t)$ | Disease-induced Mortality Rate for notified individuals* |
| $\gamma^{NN}(a,t)$ | Recovery Rate for non-notified individuals |
| $\gamma^N(a,t)$ | Recovery Rate for notified individuals |
| $\kappa(a,t)$ | Notification Rate |

* Constructed as equal to $0.15/\{1 + \exp(-0.1(a - 57.31))\}$ years$^{-1}$ as in Amaku et al. (2016b).

The solution of system (2) can be obtained with the method of characteristics (Trucco, 1965). However, for our purposes, it is better to solve the equation by following a cohort, as described in (Lopez et al., 2016).

The solution of the equation for susceptible individuals is:

$$S(a, t_0 + a) = S(0, t_0) exp\left(-\int_0^a [\lambda(s, t_0 + s) + \mu(s, t_0 + s)]ds\right). \quad (3)$$

The solution for the equation for infected individuals is:

$$I^{NN}(a, t_0 + a) = \int_0^a \lambda(s, t_0 + s) S(s, t_0 + s)$$
$$exp\left(-\int_s^a [\mu(x, t_0 + x) + \gamma^{NN}(x, t_0 + x) + \alpha^{NN}(x, t_0 + x) + \kappa(x, t_0 + x)]dx\right) ds, \quad (4)$$

$$I^N(a, t_0 + a) = \int_0^a \kappa(s, t_0 + s) I^{NN}(s, t_0 + s)$$
$$exp\left(-\int_s^a [\mu(x, t_0 + x) + \gamma^N(x, t_0 + x) + \alpha^N(x, t_0 + x)]dx\right) ds. \quad (5)$$

Finally, the equation for the removed individuals is given by:

$$R(a, t_0 + a) = \int_0^a \left(\gamma^{NN}(s, t_0 + s)I^{NN}(s, t_0 + s) + \gamma^N(s, t_0 + s)I^N(s, t_0 + s)\right)$$
$$exp\left(-\int_s^a [\mu(x, t_0 + x)]dx\right) ds. \quad (6)$$

Assuming steady state, the system (1) was solved by Amaku et al. [13] (2016) to calculate the prevalence of HCV in Brazil. The work that follows is an extension of the methods described there and its results are in accordance with the previous results for the cases where real data are available.

**Discrete time and age model**

In real life epidemics notification is discrete with the time and age units expressed in weeks, months or years. Hence, in order to apply the model to a real public health problem we discretized model (2), with time and age unit expressed in years. This discretization has to be done carefully to use the maximum advantage of the data available.

**Calculating the prevalence $I^{NN*}(A,t_i)$ and $I^{N*}(A,t_i)$**

We assume a time discrete as follows: the time $t_i$ is taken to be December 31 at 11:59 pm of year $i$; $A$ (a discrete integer variable) is the age of the individual, in whole years, at time $t_i$. So the continuous age $a$ is between $A$ and $A+1$ years, including $A$ but excluding $A+1$.

Denote by $I^{NN*}(A,t_i)$ and $I^{N*}(A,t_i)$ respectively the number of non-notified and notified individuals aged $A$ (exact age between $A$ and $A+1$) at time $t_i$. In the following the parameters and variables such as $\kappa_{A,t_i}$ and $\phi_{A,t_i}^{NN}$ mean the corresponding functions $\kappa(a, t_i)$ and $\phi^{NN}(a, t_i)$ calculated for $a \in [A, A+1)$ and $t \in (t_i-1, t_i]$.

The discretized versions of equations (4) and (5) are given by equations (7) and (8) below, which are approximations as explained in the Appendix.

$$I^{NN*}(A, t_i) = I^{NN*}(A-1, t_i-1) exp\left[-\frac{1}{2}\left(\kappa_{A-1,t_i} + \kappa_{A,t_i} + \phi_{A-1,t_i}^{NN} + \phi_{A,t_i}^{NN}\right)\right]$$
$$+ INC(A, t_i), \qquad (7)$$

where $INC(A, t_i)$ is the new HCV cases occurring between times $t_i$-1 and $t_i$ that are still alive, infectious and non-notified at time $t_i$ in the year cohort born between times $t_i$-$A$-1 and $t_i$-$A$. Here $\phi^{NN}(a,t) = \mu(a,t) + \gamma^{NN}(a,t) + \alpha^{NN}(a,t)$. In equation (7), the term

$$exp\left[-\frac{1}{2}\left(\kappa_{A-1,t_i} + \kappa_{A,t_i} + \phi_{A-1,t_i}^{NN} + \phi_{A,t_i}^{NN}\right)\right]$$

means the probability of not being removed from the non-notified class of individuals, either by natural death, disease-induced death, recovery or notification in the interval $(t_i-1,t_i]$. Equation (7) is very important because, as shown later in the paper, it allows the calculation of the true incidence from empirical data (see equation (12) below).

Recurrence (7) can be solved by well known methods and the prevalence of notified and non-notified individuals can be estimated (see equations (13) and (14) below).

Similarly, we can write:

$$I^{N*}(A, t_i) = I^{N*}(A-1, t_i-1) exp\left[-\frac{1}{2}\left(\phi^N_{A-1,t_i} + \phi^N_{A,t_i}\right)\right]$$
$$+ \int_A^{A+1} NOTIFICATION(a, [t_i-1, t_i]) da, \quad (8)$$

where $\phi^N(a,t) = \mu(a,t) + \gamma^N(a,t) + \alpha^N(a,t)$. The last term represents the notifications of HCV between times $t_i$-1 and $t_i$ of individuals in the year cohort born in $t_i$-$A$-1 to $t_i$-$A$ who are still in the notified class at time $t_i$, i.e.

$$\int_A^{A+1} \int_0^1 \kappa(a-1+x, t_i-1+x) I^{NN}(a-1+x, t_i-1+x)$$
$$exp\left[-\int_x^1 \phi^N(a-1+z, t_i-1+z) dz\right] dx\, da,$$
$$\approx \kappa\left(A+\frac{1}{2}, t_i\right) I^{NN}\left(A+\frac{1}{2}, t_i\right) \approx \kappa(A, t_i) I^{NN*}(A, t_i), \quad (9)$$

as the integration intervals are of length one and $a = A + \frac{1}{2}$, $x$=1 lies in both of them, as explained in the Appendix (equation (A5)).

Equations (7) and (8) are almost useless as written. However, in the next section, we are going to show how to solve them using the notified cases in a particular setting, namely hepatitis C in Brazil. Using the notified incidences and good guesses for the mortality rates we can calculate any desired properties of the infected population. In the next section we calculate the prevalence of the disease. The calculation presented applies to any notifiable infectious disease.

**Example of Application: Hepatitis C**

In this section we exemplify the above theory by calculating the prevalence of hepatitis C (HCV), a flaviviral infection that afflicts close to 3% of the world population (WHO, 2016), in Brazil. As mentioned in the Introduction, the great majority of infections with HCV, however, are not easily identified and, therefore, frequently non-notified. We use the Brazilian Ministry of Health official notification systems, called "Sistema Informação Nacional de Agravos de Notificação (National Information System of

Notifiable Diseases)" (SINAN, 2015), this data is denoted hereafter $SINAN(a,t)$.
Figure 1 shows the time and age variation in the reported number of HCV in Brazil.

In fact, the actual number of reported HCV infections is available only from 2000 onward. As we know from previous studies (Romano et al., 2010), HCV was introduced in Brazil in the later 1950s. We therefore constructed the number of reported with a sigmoidal decay backwards until 1932, as argued below. We used this artifice only to illustrate the model and these figures have little epidemiological significance, as argued below. We shall return to this point in the results section 4, where we explain this procedure in more detail.

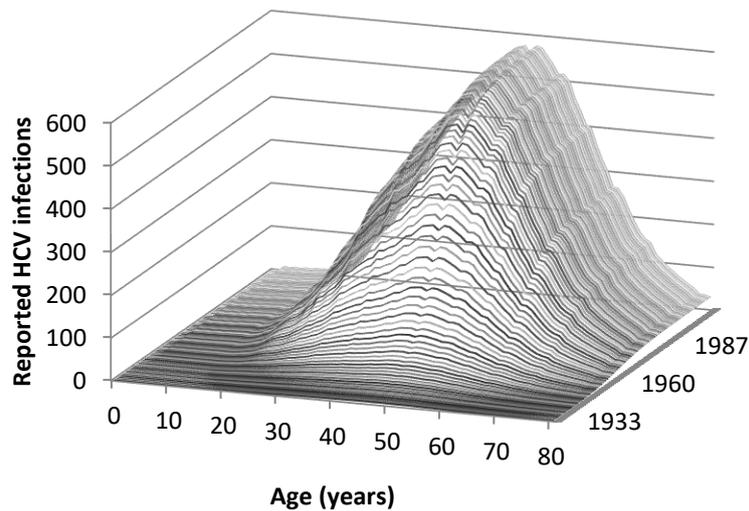

**Figure 1. Time and Age variation of the reported number of HCV infections in Brazil, artificially constructed by extrapolating backwards until 1932.**

**Estimating the total number of HCV infected individuals in Brazil**

We define $SINAN(a,t)$ to be the density with respect to age $a$ of the incidence of notified cases at time $t$. Hence $SINAN(a,t)$ has dimensions (time)$^{-2}$. As in Amaku et al. (2016) we assume that $SINAN(a,t)$ is a fraction of the density with respect to age of non-notified cases:

$$SINAN(a,t) = \kappa_{a,t} I^{NN}(a,t). \qquad (10)$$

We want to relate *SINAN(a,t)* to the SINAN database so we introduce *SINAN*(A,t_i)* to denote the number of individuals aged *A* to *A+1* at time $t_i$ who were notified to SINAN in the current year $(t_i-1, t_i]$. Now

$$SINAN^*(A, t_i) \approx \kappa_{A,t_i} I^{NN*}(A, t_i), \quad (11)$$

using equation (9).

From (7) and (11) we can write down the fundamental equation for estimating the incidence:

$$INC(A, t_i) = \frac{SINAN^*(A, t_i)}{\kappa_{A,t_i}}$$
$$- \frac{SINAN^*(A-1, t_{i-1})}{\kappa_{A-1,t_{i-1}}} \exp\left(-\frac{1}{2}\left(\kappa_{A-1,t_i} + \kappa_{A,t_i} + \phi^{NN}_{A-1,t_i} + \phi^{NN}_{A,t_i}\right)\right).$$
(12)

Note that, as observed in equation (12), the method consists of subtracting consecutive values of a diagonal of a matrix containing age in lines and time in columns. In some instances, however, it may happen that for certain ages and years the calculated incidence is negative. Our interpretation is that, for that particular age and time, the notified incidence was zero. When this happened in the actual calculation we assigned the value zero to the notification incidence.

Therefore, $I^{NN*}(A,t_i)$ can be calculated for each age and time reported as

$$I^{NN*}(A, t_i) = \sum_{j=0}^{A} INC(A-j, t_i-j)$$
$$exp\left\{-\frac{1}{2}\sum_{p=0}^{j-1}\left(\kappa_{A-1-p,t_i-p} + \kappa_{A-p,t_i-p} + \phi^{NN}_{A-1-p,t_i-p} + \phi^{NN}_{A-p,t_i-p}\right)\right\}.$$
(13)

Similarly, for $I^{N*}(A,t_i)$, we have:

$$I^{N*}(A, t_i) = \sum_{j=0}^{A} SINAN^*(A-j, t_i-j) \, exp\left\{-\frac{1}{2}\sum_{p=0}^{j-1}\left(\phi^{N}_{A-1-p,t_i-p} + \phi^{N}_{A-p,t_i-p}\right)\right\}.$$
(14)

Figure 2 shows the calculation of $INC(A, t_i)$ using equation (12) with the SINAN data as shown in Figure 1.

**The size of the Liver Transplantation Waiting List in Brazil**

It is known that a fraction of those individuals infected with HCV evolve to liver failure after many years of infection (Chaib and Massad, 2005). Let us denote those individuals diagnosed with liver failure of age $A$ at time $t_i$ as $LF(A,t_i)$. These individuals have been necessarily diagnosed with HCV and, therefore, are a fraction of the notified infected individuals $I^{N*}(A,t_i)$. It is assumed that individuals develop liver failure after a minimum time interval $\tau_{\min}$, say 10 years. From equation (8) for $I^{N*}(A,t_i)$ we obtain the equation for $LF(A,t_i)$:

$$LF(A,t_i)\Delta\tau = \sum_{\tau=\tau_{\min}}^{A} \eta_{A-\tau} I^{N*}(A-\tau, t_i-\tau)\Delta\tau\Delta A \exp\left\{-\frac{1}{2}\left[\sum_{p=0}^{\tau-1}(\phi^N_{A-1-p,t_i-p} + \phi^N_{A-p,t_i-p})\right]\Delta A\right\},$$

(15)

where $\eta_{A-\tau}$ is a discretized function that decreases from $\tau = \tau_{\min}$ up until $\tau = A$, representing the rate at which infected (and notified) individuals of age $A$-$\tau$ develop liver failure, and $\Delta \tau = \Delta A = 1$ year. Summing up over all ages we obtain the size of $LF(t_i)$, which is the total number of individuals with liver failure at time $t_i$:

$$LF(t_i)\Delta \tau = \sum_{A=A_{\min}}^{A_{\max}} \sum_{\tau=\tau_{\min}}^{A} \eta_{A-\tau} I^{N*}(A-\tau, t_i-\tau) \Delta \tau \Delta A \exp\left\{-\frac{1}{2}\left[\sum_{p=0}^{\tau-1}(\phi^N_{A-1-p,t_i-p} + \phi^N_{A-p,t_i-p})\right]\Delta A\right\}. \quad (16)$$

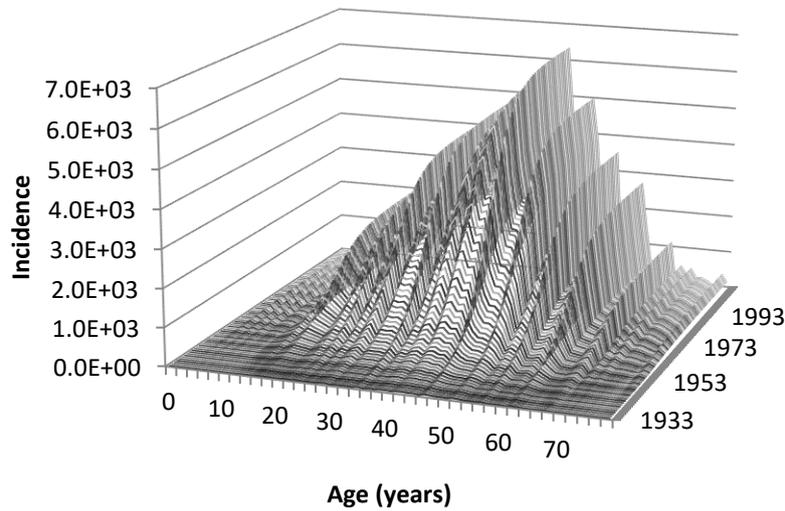

Figure 2. Calculation of $INC(A, t_i)$ from the SINAN data as shown in Figure 1.

Apart from those individuals who are transplanted (see below) $LF(t_i)$ corresponds to the Liver Transplantation Waiting List (LTWL).

Let us now rewrite equation (16) considering transplantation. Let $\psi(a,t)$ be the transplantation rate of individuals of age $a$ at time $t$. Then, equation (16) becomes

$$LTWL(t_i)\Delta \tau = \sum_{A=A_{\min}}^{A_{\max}} \sum_{\tau=\tau_{\min}}^{A} \eta_{A-\tau} I^{N*}(A-\tau, t_i-\tau) \Delta \tau \Delta A$$
$$\times \exp\left\{-\frac{1}{2}\left[\sum_{p=0}^{\tau-1}(\phi^N_{A-1-p,t_i-p} + \phi^N_{A-p,t_i-p} + \psi_{A-1-p,t_i-p} + \psi_{A-p,t_i-p})\right]\Delta A\right\}. \quad (17)$$

The number of transplants is then given by $TR(t_i)$ where

$$TR(t_i)\Delta \tau = \sum_{A=A_{\min}}^{A_{\max}} \sum_{\tau=\tau_{\min}}^{A} \psi_{A,t_i}\eta_{A-\tau} I^{N*}(A-\tau, t_i-\tau) \Delta \tau \Delta A$$
$$\times \exp\left\{-\frac{1}{2}\left[\sum_{p=1}^{\tau-1}(\phi^N_{A-1-p,t_i-p} + \phi^N_{A-p,t_i-p} + \psi_{A-1-p,t_i-p} + \psi_{A-p,t_i-p})\right]\Delta A\right\}. \quad (18)$$

We take for $\psi_{A,t_i}$ a suitably truncated bell-shaped discrete function (Chaib and Massad, 2005) with a maximum at 45 years of age for all $t_i$.

**Results and Discussion**

One of our objectives is to calculate equations (13) and (14) in order to obtain the estimated prevalence of notified and non-notified HCV infections which sum up to total prevalence. Unfortunately, the data available are restricted to the period between 2000 and 2012. In order to simulate a longer history of HCV infection in Brazil, we artificially constructed such a previous history by extrapolating backwards. First, we averaged the notified cases in the period between 2000 and 2012. Then, we fitted a sigmoidal-shaped curve representing the notified cases back for the period between 1932 and 2000. We did that for all ages such that the age distribution of notified cases was assumed fixed for all the extrapolated periods. We are well aware that HCV was probably introduced in Brazil in the 1950's and, therefore, this calculation is only an exercise to illustrate the method.

In a previous paper (Amaku et al., 2016b), this extrapolation was done differently. We assumed the disease to be in steady state until 1932. The results of this previous calculation are therefore different from the ones presented in this paper. We shall elaborate on this later. To begin with, Figure 3 shows a preliminary result on this direction. The continuous line is the total prevalence extrapolating the data as if in steady state (Amaku et al., 2016b). The sigmoid dotted line is the total prevalence calculated assuming the artificially constructed notification as explained above.

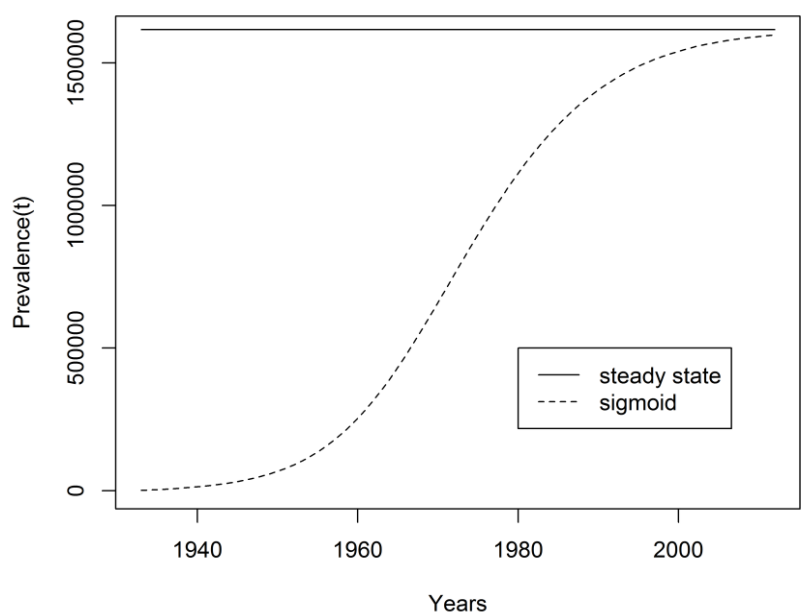

**Figure 3. Comparison of the total prevalence calculated according to Amaku et al. (2016b) (continuous line) and assuming the notification as a sigmoidal extrapolation (dotted line).**

Results of the numerical calculations are summarized in Table 2. In it we compare the prevalence in 2012 of HCV infected individuals who have been reported to SINAN until 2012 with the outcomes of the model. In Figure 4 we also compare the size of the liver transplantation waiting list according to the official figures with the outcomes of the model.

The results called first method and second method in Table 2 were obtained using the following procedure (see Amaku et al. (2016) for details). First, we assumed that the infection was in steady state from 2004 to 2012 and averaged the reported incidence. This reported incidence was extrapolated backwards until 1932. It is therefore not surprising that the published number in Amaku et al. (2016) including the third and fourth columns of Table 2 are larger than the figures obtained in this paper. The difference represents up to a certain point the state of the infection prior to 2000 and from this point of view the results seem to be consistent with what was believed about the infection in Brazil.

| Table 2. Summary of the Results | | | |
|---|---|---|---|
| **RESULTS** | Current Method | First Method of Amaku et al. (2016) | Second Method of Amaku et al. (2016) |
| **Prevalence of Notified HCV Infections** | 163,902* <br> 169,382** | 240,120[#] | 227,074[#] |
| **Prevalence of Non-Notified HCV in Brazil** | 1,433,638* <br> 1,446,771** | 1,650,100[#] | 1,632,300[#] |
| **Total Prevalence of HCV in Brazil** | 1,597,540* <br> 1,616,153** | 1,890,220[#] | 1,859,374[#] |

*Using only the official SINAN period (2000-2012) assuming zero notification incidence for all years and ages from 2000 backwards until 1932.
** Calculated from real data (2000-2012) and extending the data backwards assuming a sigmoidal decay until 1932.
[#] Taking the average number of cases reported annually to SINAN between 2004 and 2012, a period in which a steady state could be assumed.

From the results of the current method expressed in Table 2 it is possible to observe that the difference between taking into account the constructed data backwards until 1932 and the official SINAN period of 2000-2012, reflects the significant contribution of this period to both the SINAN and the total prevalence of HCV in Brazil. Note that the artificially constructed incidence will manifest itself for individuals older than 40 years.

Figure 4 shows the comparison between the actual size of the LTWL as in Chaib et al. (2014) and the result of the application of equation (17).

This paper is an attempt to provide a method to estimate the actual number of infected individuals (and other parameters related to transmission) of compulsory notifiable infectious diseases from the officially notified number of cases. Considering that, in the great majority of cases, the number of notified cases represents only a small but variable fraction of the total number of infected individuals, a reliable method of estimating the latter from the former can represent an important tool for public health policies. Notwithstanding the recognized importance of under-notification of most chronic infections, the tools to deal with this information gap proposed so far are varied and, to the best of our knowledge, there is currently no consensus about which is or are the most appropriate (Doyle, 2002; Gibbons et al., 2014; Keramarou and Evans, 2012; Rowe and Cowie, 2016; Gibney et al., 1991; Serra et al., 1999).

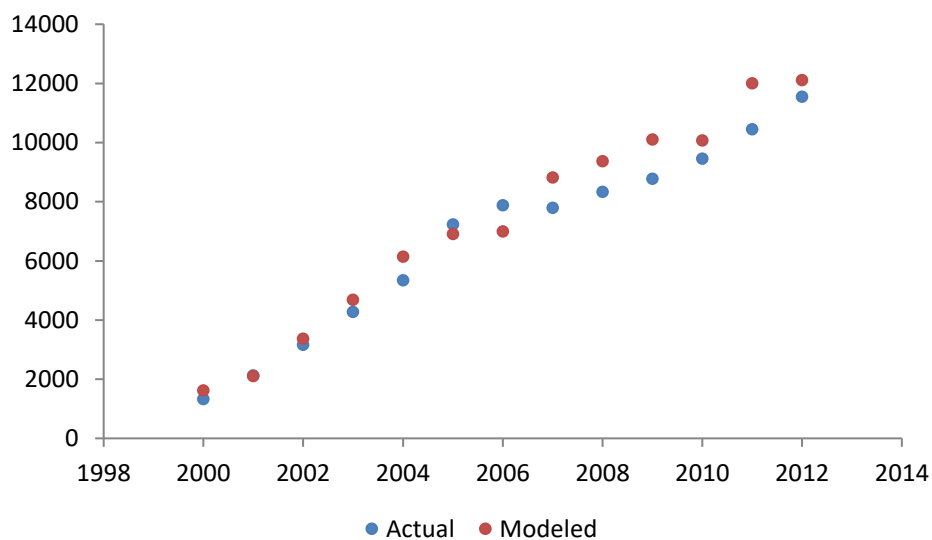

**Figure 4. Comparison between the actual size of the LTWL as in Chaib et al. (2014) and the result of the application of equation (17).**

In a previous publication (Amaku et al., 2016), a continuous time-dependent model for the estimation of the total number of HCV infected individuals in Brazil was proposed. In that paper, we assumed a steady state for the period between 2004 and 2012, and we concluded that the non-notified to notified ratio in the number of infections was about 7 to 1. The current work is an extension of that paper and we relaxed the steady state assumption. To do a calculation for individuals with age up to 80 years, we artificially extended the official notification database backwards from the year 2000 back to 1932. This artificially constructed database was intended only to illustrate the method. In addition, we discretized the variables time and age both because the notification database presents the number of cases per year and because the discrete model is easier to be implemented, both mathematically and computationally, than the continuous age and time corresponding model.

The method presented in this paper is applicable to any compulsory notifiable infectious disease provided that one has information about at least two end-points of the natural history of the disease of interest, or carrying out an alternative diagnostic test in a representative sample of the affected population. For instance, for the case of HCV, we used the number of notified cases and the size of the liver transplantation waiting list. For other diseases, in which one has only the number of notified cases, an alternative to the liver transplantation waiting list depends on the disease one is interested in. For instance, for the case of dengue in a sufficiently small region, an age-dependent seroprevalence profile of a properly designed sample of the population would be sufficient. For infections like HIV, in addition to the reported number of cases, a sample representing each group of risk should be used.

The method demonstrated to be accurate in retrieving the number of infected individuals for the case of HCV and the results are in good accordance with the previous estimations by Amaku et al. (2016).

In spite of its accuracy and simplicity, the method here presented has some important limitations that are worthwhile mentioning. Firstly, the model is data-greedy in the sense that a long time series of notified cases is necessary for the calculations. Secondly, the model has a large number of parameters whose values are not known with any precision for the great majority of cases. For example, as the model deals with long time series, demographic parameters such as the natural mortality rate are crucial for the calculations.

Notwithstanding those limitations, the model has the advantage that it can predict quantities that can be iteratively used to improve it. For instance, for HCV the model allows the calculation of the proportion of individuals that have the infection for $\tau$ years, that is the age of infection. If this can be checked from information from patients (e.g., blood transfusion time), the model can be improved immediately. This is thoroughly explained in Amaku et al. (2016).

**Conclusions**

We can conclude that the model proposed in this paper can be useful for estimation of the actual magnitude of endemic states of infectious diseases, particularly for those where the number of notified cases is only the tip of the iceberg. In addition, the method can be applied to other situations, such as the well known underreported incidence of criminality (for example rape), among others.

**Acknowledgements**

This work was partially funded by LIM01-HCFMUSP, CNPq, Brazilian Ministry of Health (Grant TED 27/2015) and FAPESP. DG is grateful to the Leverhulme Trust for support from a Leverhulme Research Fellowship (RF-2015-88) and the British Council, Malaysia for funding from the Dengue Tech Challenge (Application Reference DTC 16022). EM and DG are grateful to the Science Without Borders Program for a Special Visiting Fellowship (CNPq grant 30098/2014-7).

**Declaration of Interest**

None

**Appendix**

In this Appendix, we deduce the equation (7) from the main text. Let us define the function $I^{NN}(a+x, t+x)$, which is a function that expresses the evolution of a cohort. Then

$$\frac{d}{dx}[I^{NN}(a+x,t+x)] = \lambda(a+x,t+x)\,S(a+x,t+x)$$
$$-[\kappa(a+x,t+x) + \phi^{NN}(a+x,t+x)]I^{NN}(a+x,t+x),$$
(A1)

where $\phi^{NN}(a+x,t+x) = \mu(a+x,t+x) + \gamma^{NN}(a+x,t+x) + \alpha^{NN}(a+x,t+x)$.

Multiplying both sides by $exp[\int_0^x (\kappa(a+z,t+z) + \phi^{NN}(a+z,t+z))dz]$, we have

$$\frac{d}{dx}\left[exp\left[\int_0^x (\kappa(a+z,t+z) + \phi^{NN}(a+z,t+z))dz\right] I^{NN}(a+x,t+x)\right] =$$

$$\lambda(a+x,t+x) S(a+x,t+x) exp\left[\int_0^x (\kappa(a+z,t+z) + \phi^{NN}(a+z,t+z))dz\right].$$

(A2)

So integrating we deduce that

$$I^{NN}(a,t) = I^{NN}(a-1,t-1) \tag{A3}$$

$$exp\left[-\int_0^1 \{\kappa(a-1+z,t-1+z) + \phi^{NN}(a-1+z,t-1+z)\}dz\right]$$

$$+\int_0^1 \lambda(a-1+x,t-1+x)S(a-1+x,t-1+x)$$

$$exp\left[-\int_x^1 \{\kappa(a-1+z,t-1+z) + \phi^{NN}(a-1+z,t-1+z)\}dz\right]dx.$$

The first term corresponds to non-notified individuals ages *a*-1 at time *t*-1 who remain infectious and non-notified at time *t* (when their age is *a*). The second term which we denote

*INCIDENCE(a,(t-1,t])*

is the density with respect to age *a* of the incidence of HCV in the cohort of individuals born at time *t-a* which occurs in the time interval (*t*-1,*t*) and is still infectious and not notified at time *t*.

Now, $I^{NN*}(A,t_i)$, the absolute number of infectious non-notified individuals of age in the interval [*A*,*A*+1) at time $t_i$,

$$= \int_A^{A+1} I^{NN}(a,t_i)da, \tag{A4}$$

$$\approx I^{NN}\left(A+\frac{1}{2},t_i\right) \tag{A5}$$

taking the midpoint as an approximation.

Now from (A3) and (A4)

$I^{NN*}(A,t_i) = \int_A^{A+1} I^{NN}(a-1,t_i-1)$

$$exp\left[-\int_0^1 \{\kappa(a-1+z,t_i-1+z) + \phi^{NN}(a-1+z,t_i-1+z)\}dz\right]da$$

$$+\int_A^{A+1} INCIDENCE(a, [t_i - 1, t_i])da. \tag{A6}$$

The last term in (A6), which we shall denote $INC(A, t_i)$, represents the incidence between times $t_i$-1 and $t_i$ of HCV that is still infectious and not notified at time $t_i$, in the cohort born between times $t_i$-$A$-1 and $t_i$-$A$. In the first term in (A6) again for the $a$-integration we take $a = A + \frac{1}{2}$ as an approximation.

$$I^{NN*}(A, t_i) = I^{NN}\left(A - \frac{1}{2}, t_i - 1\right)$$
$$exp\left[-\int_0^1 \left\{\kappa\left(A - \frac{1}{2} + z, t_i - 1 + z\right) + \phi^{NN}\left(A - \frac{1}{2} + z, t_i - 1 + z\right)\right\}dz\right]$$
$$+ INC(A, t_i).$$

$$= I^{NN}\left(A - \frac{1}{2}, t_i - 1\right)$$
$$exp\left[-\int_0^1 \left\{\kappa\left(A - \frac{1}{2} + z, t_i\right) + \phi^{NN}\left(A - \frac{1}{2} + z, t_i\right)\right\}dz\right] + INC(A, t_i),$$

as $\kappa\left(A - \frac{1}{2} + z, t\right) + \phi^{NN}\left(A - \frac{1}{2} + z, t\right)$ is the same for $t \in (t_i - 1, t_i]$.

$$\approx I^{NN*}(A - 1, t_i - 1)exp\left[-\frac{1}{2}\left(\kappa_{A-1,t_i} + \kappa_{A,t_i} + \phi^{NN}{}_{A-1,t_i} + \phi^{NN}{}_{A,t_i}\right)\right]$$
$$+ INC(A, t_i).$$